\documentclass[final,5p,times,twocolumn]{elsarticle}

\usepackage[english]{babel}
\usepackage{graphicx}
\usepackage{amssymb}
\usepackage{amsmath}
\usepackage{amsfonts}
\usepackage{color}
\usepackage{soul}
\usepackage[update,prepend]{epstopdf}

\journal{Current Opinion in Colloid \& Interface Science}

\begin{document}

\begin{frontmatter}

\title{Inverse patchy colloids: synthesis, modeling and self-organization}

\author[label1,label2]{Emanuela Bianchi}
\ead{emanuela.bianchi@tuwien.ac.at}
\author[label3]{Peter D.~J. van Oostrum}
\author[label1]{Christos N. Likos}
\author[label2,label4]{Gerhard Kahl}

\address[label1]{Faculty of Physics, University of of Vienna, Boltzmanngasse 5, A-1090, Vienna, Austria}
\address[label2]{Institute for Theoretical Physics, Technische Universit{\"a}t Wien, Wiedner Hauptstra{\ss}e 8-10, A-1040, Vienna, Austria}
\address[label3]{Institute for Biologically Inspired Materials, University of Natural Resources and Life Sciences, Muthgasse 11/II, A-1190 Vienna, Austria}
\address[label4]{Center for Computational Materials Science (CMS), Technische Universit{\"a}t Wien, Wiedner Hauptstra{\ss}e 8-10, A-1040, Vienna, Austria}

\begin{abstract}
Inverse patchy colloids are nano- to micro-scale particles with a surface divided into differently charged regions. This class of colloids combines directional, selective bonding with a relatively simple particle design: owing to the competitive interplay between the orientation-dependent attraction and repulsion -- induced by the interactions between like/oppositely charged areas --  experimentally accessible surface patterns are complex enough to favor the stabilization of specific structures of interest. Most important, the behavior of heterogeneously charged units can be ideally controlled by means of external parameters, such as the pH and the salt concentration. We present a concise review about this class of systems, spanning the range from the synthesis of model inverse patchy particles to their self-assembly, covering their coarse-grained modeling and the related numerical/analytical treatments.

\end{abstract}

\end{frontmatter}

\section{Introduction}\label{sec:introduction}

In the last fifteen years, patchy colloidal particles have emerged as one of the most promising  solutions to the problem of designing self-assembling target structures with specific properties at the micro- and nano-scale level~\cite{Glotz_2004}. Patchy particles are indeed able to guarantee a fine control over the features of the equilibrium, ordered as well as disordered, phases due to the anisotropy of their interaction patterns: the orientational and possibly selective bonding mechanism mediated by the patches controls the symmetries of the resulting structures~\cite{Zhang_2004,patchyrevexp,patchyrevtheo}. 
Many synthesis techniques have been developed to produce colloids with chemically or physically distinct surface areas so that directional and possibly site-specific bonds between the particles can be induced: non-spherical colloidal clusters functionalized with DNA strands~\cite{Pine_2012} and spherical colloids with metallic~\cite{Granick_2011,Kretz_2013} or polymeric patches~\cite{Kumacheva_2016} are just some examples of the available synthesis advancements. The spontaneous assembly of patchy colloids has been experimentally observed for instance in the site-specific aggregation of finite colloidal clusters~\cite{Pine_2012} or in the formation of an extended, two-dimensional crystal, known as kagome lattice~\cite{Kagomeexp_2011};  at the same time, numerical and theoretical investigations in the bulk have pointed out a surprising multitude of phases that significantly extends the many possibilities already offered by isotropically interacting colloids~\cite{patchyrevtheo}. 

Recently, a new direction of investigation in the field of patchy systems has been put forward:
soft patchy units have been introduced, combining directional interactions and low bonding valence with soft interactions and mobile patches. Examples of soft patchy particles are colloids or emulsion droplets with mobile DNA linkers~\cite{Feng2013,Brujic_2013}, DNA nano-stars~\cite{Biffi_nanostars_2013} and  polymer-based aggregates~\cite{capone:prl_2012}. The particular features of these patchy units are the possible fluctuations in the positions and in the size of the patches, resulting in an enhanced bond flexibility with respect to conventional patchy colloids. So far, soft patchy particles have shown a general tendency to stabilize disordered phases~\cite{Rovigatti_acs_2014,smallenburg_2013,bianchisoft_2015}.

A further direction of investigation in the field of patchy particles has started with the introduction of particles with heterogeneous surface charge. These units can be described as patchy colloids with differently charged surface regions and have been referred to as inverse patchy colloids (IPCs)~\cite{bianchi:2011}:  while conventional patchy particles are realized by adding attractive patches on the surface of otherwise repulsive particles, IPCs have patches that repel each other, while they attract the rest of the colloidal surface that is free of patches.  The feature that the respective patches of two IPCs repel each other while they attract the bare regions of the colloid represents a striking difference with respect to conventional patchy colloids -- thus the specification ``inverse patchy''. The interplay of attraction and repulsion of oppositely and like charged regions leads to a complex effective potential between IPCs, which depends on both the separation of the two interacting particles and their mutual orientation.

Although it seems obvious that heterogeneous surface charges play an important  role in a wide range of systems, so far there have been only limited  experimental studies dedicated to the self assembly of IPCs. 
One of the first experimental papers demonstrating the possibility to make colloids interact as heterogeneously charged units showed that charged Janus, overall neutral, zwitterionic particles, can indeed act as charge dipoles, forming strings~\cite{Cayre2003}. The importance of the electrostatic screening length to the interactions between heterogeneously charged colloids was elegantly demonstrated with charged Janus spheres, that where shown to form clusters due to the interactions between oppositely charged hemispheres in the presence of salt to shorten the interaction length~\cite{Hong2006}.
Charged Janus particles are also created by attaching charged polymeric particles to a liquid-liquid interface: if small oppositely charged particles are dispersed in one of the liquids and if the screening length is much smaller than the big particles' size, then the small particles attach to the big particles, thus inverting the surface charge of the involved hemisphere~\cite{Sabapathy2016}. 
Heterogeneously charged complexes can also emerge by aggregation of oppositely charged particles of different sizes~\cite{Demirrs2015}, the shape of these aggregates being controlled by the size ratio among the charged colloids as well as by the electrostatic screening length. Incidentally, the distribution of the satellite particles on the surface of the larger colloidal particles -- within the same aggregate -- could also be manipulated by the application of external electric fields.
Finally and most importantly in the context of the present review, spherical particles with two positively charged polar patches and a negatively charged equatorial belt have been recently realized by {\it ad hoc} surface modifications~\cite{peterjpcm_2015}: this study is explicitly dedicated to the synthesis and the assembly of IPCs in two-dimensions, laying the basis for a vast variety of experimental scenarios. 

It is worth noting that, besides their application in materials science, colloids with differently charged surface regions can be also considered as models for more complex biological systems: proteins and virus capsids are known to exhibit heterogenous surface charges~\cite{bromovirus,lysozyme,Podgornik_2012} and their complex behavior is affected by the anisotropic nature of their effective interactions~\cite{Schreiber2014,Lund2012,Oskolkova2015}.

The first elaborate model for particles with heterogeneously charged surfaces, for which the term IPC was coined, was put forward some five years ago in Ref.~\cite{bianchi:2011}. The model was inspired by complex units resulting from the condensation of two polyelectrolyte stars on the surface of an oppositely charged colloid~\cite{Likos_2008} (for an artistic view see the cover of Ref.~\cite{patchyrevtheo}) and it was later used to describe simpler experimental colloids with a charged surface pattern~\cite{peterjpcm_2015}. In Ref.~\cite{bianchi:2011} an IPC was modeled as a charged sphere decorated with two oppositely charged regions of finite extent (termed patches), located at the respective poles of the colloid (see Figure~\ref{fig:model}); in Ref.~\cite{bianchi:2015} this original model was recently generalized to characterize colloids with possibly richer surface patterns, such as IPCs with two differently charged or sized patches as well as IPCs with more than two patches. This model has been used to numerically investigate the collective properties of this class of particles, such as the gas-liquid phase separation, the formation of gel-like structures or the stability of new emerging crystals.

In the subsequent years, the idea to study particles with heterogeneously charged surfaces proliferated within the community. In Ref.~\cite{Cruz_2016} a charged Janus model was developed, represented by a charged, spherical colloid, decorated with one single oppositely charged patch. The (pairwise) interaction between these particles is considerably simpler than the one derived for the model put forward in Ref.~\cite{bianchi:2011}: the radial part consists of a soft, steric repulsion plus an electrostatic term (for the differently charged regions), treated within the Debye-H\"uckel theory; the latter contribution is multiplied by a distance- and orientation-dependent prefactor which generalizes ideas of the seminal Kern-Frenkel model for conventional patchy particles~\cite{KernModel_2003}.  This model has been used to study the aggregation behavior of such units depending on their patch size~\cite{Cruz_2016}.
Again along the Debye-H\"uckel lines, Ref.~\cite{Hiero_2016} proposes a multipolar expansion for the interaction potential between charged Janus colloids: in this work, minimum-energy clusters relative to different orders of the multipolar expansion are compared to experimental clusters, thus showing how such an order affects the resulting clusters. 
It is worth noting here a peculiarity of the Debye-H{\"u}ckel multipolar expansion: each standard multipole carries also all higher multipoles, as first shown in Ref.~\cite{Trizac_2000} and later elaborated in Ref.~\cite{Ramirez_2008}.
In the context of globular proteins, a set of charged patchy particle models, introduced in Ref.~\cite{Yigit15a}, can be considered as IPC-related particles: here the charged patches are generated by distributing -- following well-defined rules -- a few hundred charged beads on a spherical surface. This set of units has been subsequently used to study their adsorption on a polyelectrolyte brush layer~\cite{Yigit15b}. 
Zwitterionic and protein-like units have been also modeled in Ref.~\cite{Blanco} to elucidate the properties of the fluid phase upon small changes of the system parameters, such as the strength of the interactions or the arrangement of charged patches. 



A highly sophisticated model for particles with heterogeneously charged surfaces has been put forward in~\cite{Podgornik_2013}, which considers homogeneously charged shells, decorated with an arbitrary pattern of opposite charges within the Debye-H\"uckel approximation; particular emphasis is put on heterogeneous icosahedral, octahedral, and tetrahedral charge decoration. With its complexity, this versatile and rather general model is able to describe, for instance, virus capsids. However, the emerging analytical expressions for the
interaction free energy between two such shells seems to be too complex to be amenable to simulations and/or theoretical frameworks.

For completeness one should also mention in this context colloidal particles whose surfaces are not explicitly decorated with charges but rather carry an explicit (dipolar, quadrupolar, etc.) moment, which might originate from an inhomogeneous surface charge decoration.  We limit ourselves to mostly electric multipoles and we specifically refer to Ref.~\cite{Schmiedle2013}, where particles with either one or two, oppositely oriented dipole moments are considered for the assembly of two-dimensional colloidal networks induced by a uni-axial external field, or to Ref.~\cite{Bharti2016} where the authors study the self-assembly scenarios of colloidal particles carrying two independent dipolar moments under the effect of concurrent fields.

In this review, we summarize the achievements reached so far for the most studied IPC systems. The paper is organized as follows:  Section~\ref{sec:experiments} describes recent progress in synthesizing model IPCs; Section \ref{sec:model} focuses on the original coarse-grained description of IPCs; Section~\ref{sec:simulations} summarizes the numerical methods that have been used to investigate the collective behavior of the coarse-grained IPCs; Section~\ref{sec:theory} briefly describes the theoretical approaches that can be used to study the structural and thermodynamic properties of these model systems; Section~\ref{sec:results} gives an overview of the emerging results in experiments, theory and simulations; the paper is closed with an outlook to possible future research on IPCs.

\section{Synthesis}\label{sec:experiments}

Recently, a novel method that does not rely on clean-room facilities and that is easily scalable has allowed to modify the surface of negatively charged silica particles by creating two positively charged polar regions (see panel (a) of Figure~\ref{fig:expsim})~\cite{peterjpcm_2015}.
The particles were first ordered into a two-dimensional hexagonal crystal on top of an agarose gel thanks to the capillary forces generated upon evaporation of the dispersion medium. Then, polystyrene dissolved in diethylether was applied to the two-dimensional crystal, thus filling the space between the silica particles. Upon evaporation of  the diethylether, a polystyrene membrane formed, from which the particles protruded on both sides. This membrane was used as a template for the subsequent modification of the polar regions by amination and labeling with the cationic fluorescent dye Rhodamine B isothiocyanate. Electrophoresis measurements showed that the positive surface charge of the polar patches can be tuned via the pH of the solution in the range from pH $\sim$ 5 to 9, while the negative charge of the equatorial regions was essentially unaltered. 
The formation of large two-dimensional gel-like structures when particles sedimented on top of the bottom glass (see panels (b) of Figure~\ref{fig:expsim}) suggests that the interactions between the colloids are indeed dominated by anisotropy. Although the particles were monodisperse in size, the distribution of patch opening angles was probably too broad to allow the formation of the characteristic microcrystalline gels that where reported in simulations (see Section~\ref{sec:results})~\cite{bianchi:2d2013,bianchi:2d2014}.

In order to study the assembly of IPCs in a controlled manner, it is important to narrow the 
size distribution of the patches further down. Moreover, to facilitate the microscopic study of any self-assembled 
structure, it is advantageous that the particles are dispersed in a refractive index matching solvent~\cite{VanBlaaderen1992}. Additionally, for self-assembly to take place in the bulk, a relatively large amount of particles is needed that should also be density matched to suppress sedimentation or creaming towards a hard wall. This effect can also be achieved by working with significantly smaller particles: the gravitational length has to be large enough to make sedimentation irrelevant. However, for too small particles the visualization of their orientation is impossible. In order to also visualize the orientational order in any self-assembled structure with light microscopy, the patches have to be individually discernible, which means that the patches on a single particle have to be at least one half to one micrometer apart, providing a clear lower limit for the particle size.
The experimental efforts of developing a density and refractive index matched colloidal dispersion with control over the acidity and electrostatic  screening length are justified by the tantalizing promise of observing completely novel structures (see Section~\ref{sec:results})~\cite{ismene,silvanonanoscale}.

Several alternative methods to make patchy particles with two possibly charged polar patches have been proposed in literature. The most commonly used method to produce patches is still glancing angle deposition~\cite{Pawar2009} 
of for instance a gold layer on particles in a two-dimensional array followed by a thiol functionalisation of the patch. 
This approach can be used to make conventional hydrophobic patches~\cite{Kagomeexp_2011} as well as charged 
regions~\cite{Hong2006}. The possibilities to control the relative orientation and size of the patches with 
glancing angle deposition techniques are rather extensive: one can combine the use of different angles of deposition, the casting of shadows in ordered particles arrays and controlled etching of deposited patches~\cite{Granick_2011,Pawar2009}. Glancing angle deposition can be applied to make patches both on oxide particles like silica and on polymeric particles; the latter can be density matched. For the self assembly of structures in three dimensions, the weight of the patches is probably not a problem due to the limited amount of material deposited, nonetheless the light absorption by  the patch metal might be hampering the real space observation of any three-dimensional structures. 

A template-based method that has been reported to yield particles with charged polar patches~\cite{Lin2010} relies on the balance between the temperature dependence of the mechanical properties of electrospun polymer microfibers and their affinity with colloidal particles, which is used to control the partial coverage of the particle surface. This method gives control over the patch sizes but it depends critically on the polymer fiber composition and on the fact that not all particle materials are suitable for exposure to high temperatures. In particular, the silica particles of Ref.~\cite{Lin2010} do indeed withstand the used temperatures without any problems but they cannot be density matched for self-assembly studies in three dimensions; in contrast polymeric particles with a lower, possibly matchable density, tend to deform when heated. 

Finally, a conceptually simple manner to make particles with one or two polar patches is based on contact printing \cite{Cayre2003,Jiang2009,Seidel2016}. Via the stiffness of the stamp, the applied pressure and the amount of ``ink" used, some control can be gained over the size of the produced patches. This technique does allow for the usage of polymeric particles that can be density matched for self-assembly studies in three dimensions.

Currently, an extension of the method reported in Ref.~\cite{peterjpcm_2015} is being developed to create density and refractive index matchable polymeric IPCs. In order to deposit a template in between the particles, a solvent needs to be used that does not dissolve or swell them; on the other hand, the template material should be selectively soluble after patch modification. The agar gel used in Ref.~\cite{peterjpcm_2015} was inhomogeneous in stiffness, leading to patch size polydispersity, but can be replaced with more homogeneous substrates to improve the monodispersity of the patch sizes. After optimization of the protocol we expect to be able to study the self-assembly behavior of IPCs both in the bulk and under confinement.

\section{Modelling}\label{sec:model}
The coarse-grained description of the pair potential between IPCs is derived by combining a mean field theoretical description with  a simple patchy model.  The coarse-grained modeling was originally developed to describe IPCs carrying two polar patches with identical size and charge~\cite{bianchi:2011} and was subsequently extended to richer surface morphologies, in particular it was thoroughly carried out for two patches with different charge and/or size~\cite{bianchi:2015}. 

The mean field theoretical description mimics each particle as a charged, dielectric sphere surrounded by a dielectric solvent; on its surface it is decorated with an arbitrary number of oppositely charged regions of finite extent. Surrounding co- and counter-ions are treated within a mean-field type concept. With standard methods of electrostatics the potential generated by such a particle can be readily calculated within a Debye-H\"uckel type theory. This approach allows to determine the potential generated by one single colloid dispersed in a liquid solvent~\cite{debyehueckel}: the potentials inside (where a discrete charge distribution reproduces the symmetries of the surface pattern) and outside the colloid (where co- and counter-ions are present) are calculated separately as expansions in terms of spherical harmonics and are then linked together by electrostatic boundary conditions; the resulting set of linear equations for the yet undetermined expansion coefficients is solved, either analytically~\cite{bianchi:2011} (relying on approximating the ratio between spherical Bessel functions of consecutive orders with a Yukawa-like term) or numerically~\cite{Likos_2004,bianchi:2015}.  The analytical approximation, that can be carried out only under high screening conditions, leads to a Yukawa-like functional form.
Once the single particle potential is known, the effective interaction energy between two IPCs is determined: first the potential energy due to the presence of an IPC in the screened electrostatic field generated by another IPC is calculated and is then symmetrized; finally, the total interaction energy for a given particle-particle configuration is obtained as the average value over the two contributions normalized by the minimum of the attraction. The resulting pair potential is the sum of the screened Coulomb potentials generated by the source sphere and centered at the positions of the effective charges inside the probe sphere; in the sum, each contribution is multiplied by the magnitude of the corresponding effective charge at that point.
The resulting pair potential  depends on the distance between the two colloids and on their mutual orientation.
It is worth noting that this approach is very general and is always reliable as long as the centers of charge of the different surface areas do not lie outside the colloidal particle. The advantage of the Yukawa-like approximation consists in providing an analytic expression that is immediately recognizable as a generalized DLVO potential~\cite{dlvo}; on the other hand, the numerical approach can be kept as general as possible so that it can be applied to a wide range of different surface patterns.

\begin{figure}[htbp]
	\begin{center}
	\includegraphics[width=0.45\textwidth]{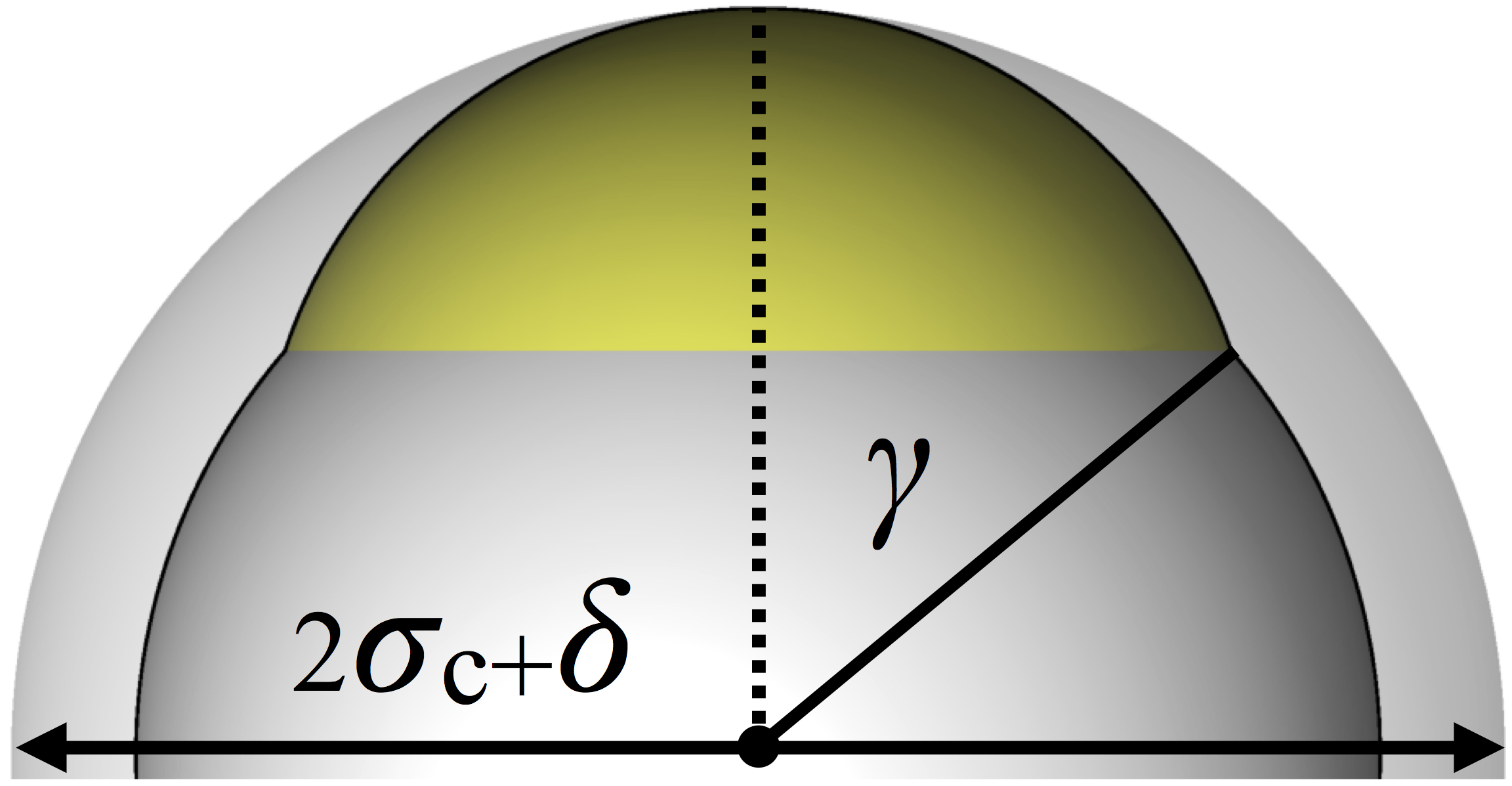} 
	\end{center}
	\caption{In the schematic representation of an IPC, the particle core is depicted in dark grey, while patches are colored in yellow. Note that IPCs are spherical units: the yellow caps represent the interaction sphere of the patches, while the light gray halo features the interaction sphere of the bare colloid. The characteristic parameters are: $\sigma_{\rm c}$, the radius of the colloids, $\delta$, the particle interaction range, and $\gamma$, the patch opening angle, as labeled. In this model~\cite{bianchi:2011}, $\sigma_{\rm c}$ is the unit of length,  while $\gamma$ and $\delta$ are determined once the position and the interaction range of the out-of-center sites are chosen. In the experimental realization~\cite{peterjpcm_2015}, $\sigma_{\rm c}$ and $\gamma$ are determined at the synthesis level, while $\delta$ depends on the overall properties of the system, such as the screening conditions, the charges of the different surface areas, the pH, the salt concentration and the dielectric constant of the surrounding medium.}
	\label{fig:model}
\end{figure}

The simple patchy model has three sets of independent parameters: the interaction ranges of the different surface regions, their surface extents and their interaction strengths (see Figure~\ref{fig:model}). The model features a hard spherical particle carrying $n$ interaction sites placed in a well-defined geometry inside the colloid. As a consequence, the corresponding interaction spheres extend partially outside the hard core particle, defining in this way the patches.
Since the characteristic interaction distances are determined by the electrostatic screening of the surrounding solvent, all entities of the colloid are assumed to have the same interaction range irrespective of the surface regions involved in the interaction. The patch size is ideally determined by the corresponding feature of experimentally synthesized particles. Finally, the energy parameters of the model are related to the charges involved in the interactions. These charges are responsible for the ratio between the attractive and repulsive contributions to the pair energy associated to the different (patch/patch, patch/bare and bare/bare) interactions. The specific form of the pair potential is based on the postulate that each of these contributions can be factorized into an energy strength and a geometrical weight factor, the latter one being given by the distance dependent overlap volume of the involved interaction spheres (each of them representing either the bare particle or a patch)~\cite{bianchi:2011}.

The Debye-H{\"u}ckel potential is subsequently used to provide the simple patchy model with parameters that are directly related to the physical quantities of the underlying microscopic system. The resulting coarse-grained description fully preserves the rotational symmetry of the IPC and it is advantageous in many-body simulations.



\section{Simulation approaches}\label{sec:simulations}

Monte Carlo (MC) simulations (performed in different types of ensembles) can be applied in a rather straightforward manner to IPC systems: in addition to the conventional translational trial moves, also trial rotations have to be applied to the particles~\cite{frenkelsmit}.  The coarse grained model described in Section~\ref{sec:model} allows for a fast evaluation of the pair interaction energy at each MC step: the calculation of the model interaction between two particles is the sum over simple products of geometric and energy factors, the first ones being orientation- and distance-dependent contributions and the latter ones being numerical coefficients, both associated to the different interactions that characterize the IPCs. For the calculation of the phase diagrams of IPC systems, free energy calculations can be performed via thermodynamic integration~\cite{NoyaEinst}: for non-spherical potentials, the Hamiltonian of the interacting Einstein crystal must include the original interaction potential of the model, an orientational and a translational field. The candidate crystal structures for such calculations were usually selected by an evolutionary approach~\cite{doppelbauer:sm2012}. The free energy of the fluid phase can be evaluated by straightforward thermodynamic integration~\cite{frenkelsmit} using as reference system either the ideal gas (i.e., integrating from very low densities) or the hard-sphere fluid (i.e., integrating from very high temperatures at which the IPCs behave as hard spheres). Starting from one known coexistence point, the entire coexistence line can be traced by using the Gibbs-Duhem method~\cite{kofkeMP}. 
MC simulations have been used to derive the collective behavior of several IPCs systems (either under confinement or in the bulk), such as network formation, gas-liquid phase separation and the stability of crystalline {\it versus} fluid phases.

If the calculation of dynamic properties (such as the mean-squared displacement, the velocity autocorrelation function, or the intermediate scattering functions) is required, Molecular Dynamics (MD) simulations are the method of choice. The most pragmatic access is to consider IPCs as rigid molecules, each of them composed by a suitable number of spherical interaction units. Also here, one can take benefit of well-documented simulation methods~\cite{frenkelsmit}: the decomposition of the particle motion into a translatory center-of-mass motion plus an additional spatial rotation represents one option. In contrast, an alternative approach has been used so far for IPCs (for a detailed justification of this choice we refer to Ref.~\cite{silvano_phdthesis}): for reasons of computational simplicity, a constrained dynamics was applied, where the dynamics of all interaction sites of a particle was computed, while imposing both the rigidity of the particle and the fixed distance between pairs of sites via Lagrange multipliers in the equations-of-motion. In case that the interaction sites form a linear arrangement (as it is the case in a symmetric, two patch IPC) particular care has to be taken since the traditional velocity Verlet integrator (RATTLE)~\cite{And83} becomes singular. To overcome this problem, a scheme introduced by Ciccotti {\it et al.}  \cite{Cic82} has been used, which explicitly reduces the number of degrees-of-freedom of rigid bodies~\cite{silvano_phdthesis}: the two patches were treated as two particles separated by a fixed distance and move according to effective forces that include both the inertia of the central particle and the forces acting on it. As a consequence of these effective forces, the central particle automatically satisfies the constraint of being in the middle point of the line joining the two patches and its trajectory is automatically provided by the knowledge of the patch trajectories. This approach reduces the number of equations of motion from 18$N$ to 12$N$ for a system of $N$ IPCs. 
MD simulations were used to investigate the self-assembly and dynamic properties of IPCs systems in the bulk~\cite{silvano_phdthesis}. 

\section{Theoretical approaches}\label{sec:theory}

While the extension of simulation-based methods to the typical interactions of IPCs might be cumbersome, but essentially straightforward, the situation is significantly more delicate when it comes to extend existing theoretical frameworks in an appropriate manner to IPC-like particles.

One of the most promising attempts in this direction is based on Wertheim's theory to describe properties of associating fluids~\cite{Werth1,Werth2}. Including ideas put forward by Smith and Nezbeda~\cite{Smi81}, Wertheim extended his original framework~\cite{Wer88}, thus providing a versatile and appropriate tool to describe the structural and thermodynamic properties of patchy particles in general.  The basic relation of Wertheim's formalism is a generalization of the original Ornstein-Zernike (OZ) equation (see, e.g.,~\cite{hansen}), where the involved correlation functions are expanded in terms of so-called bonded correlation functions, each one related to subgroups of particles in specific bonding states. These correlation functions then fulfill the so-called multi-density OZ equation, which is complemented by a suitably generalized closure relation, such as the associative Percus-Yevick or the associative hypernetted-chain relation. The ensuing integral-equations can be solved with a reasonable amount of computational effort. This formalism has been extended and applied to the case of IPCs by Kalyuzhnyi and co-workers~\cite{Yura2015-1,silvanojpcm} with partly analytical solutions for simplified versions of the model~\cite{Yura2015-2}. An alternative OZ-based approach, put forward for patchy particles is based on an expansion of the correlation functions~\cite{Gia14} and can possibly be extended to inverse patchy systems as well. The structural and thermodynamic properties of IPCs can also be obtained via perturbative schemes: we refer to a suitable extension of the Barker-Henderson second-order thermodynamic perturbation theory, originally put forward by Wertheim in Ref.~\cite{Wer87} and applied for the one-patch Kern-Frenkel model in Ref.~\cite{Goeg12}; this approach was recently extended to IPCs decorated with an arbitrary number of patches~\cite{Step16}. 
From the comparison between simulation data and the aforementioned theoretical descriptions, it emerged that the multi-density OZ equation approach with the Percus-Yevick closure guarantees a higher accuracy of the results and a better convergence ability, especially at low temperatures~\cite{Yura2015-1,silvanojpcm}. 

An alternative and possibly promising route to study the thermodynamic properties of particles with decorated surfaces has been put forward by Shin and Schweizer~\cite{Shin14}. While the previous approaches have been developed mainly to investigate low density phases, the description developed in Ref.~\cite{Shin14} is particularly suitable to investigate high density states, such as crystalline ones. The approach is based on a suitable extension of the so-called self-consistent phonon theory to particles with heterogeneously decorated surfaces. This rather general framework is able to include coupled translational and orientational entropic and enthalpic contributions to the thermodynamic potentials. Even though this approach was formulated for a simple one-patch Kern-Frenkel model in two dimensions, this framework definitely has the potential to be extended to an arbitrary number of patches and three dimensional systems, possibly also for IPC systems; activities in this direction are currently being pursued~\cite{Kal17}.

\section{Trends in the collective behavior}\label{sec:results}
The capacity of IPCs to form strongly directional, selective bonds is the reason why these particles are able to create very characteristic bonding patterns. So far, mostly IPCs with two identical polar patches have been thoroughly investigated: the common feature of their bonding patterns is the direct contact of a polar region of one particle with the oppositely charged bare, equatorial area of the other interacting
particle; the orientation of the bonding pattern depending mainly on the patch size. This pattern forms rather rapidly and spontaneously, and is characterized by a strong self-healing capacity. 
Numerical investigations on coarse-grained IPCs with two identical polar patches have shown that an emerging feature of these systems is the formation of planar aggregates either as monolayers close to a charged substrate~\cite{bianchi:2d2013,bianchi:2d2014} or as bulk equilibrium phases~\cite{ismene,evaemanJPCM,silvanonanoscale,silvano_phdthesis}:
depending on the imposed confinement and/or the available space, the described bonding pattern induces a zoo of equilibrium configurations documented in the literature and summarized in the following. 

\begin{figure}[ht]
\begin{center} 
\includegraphics[width=0.5\textwidth]{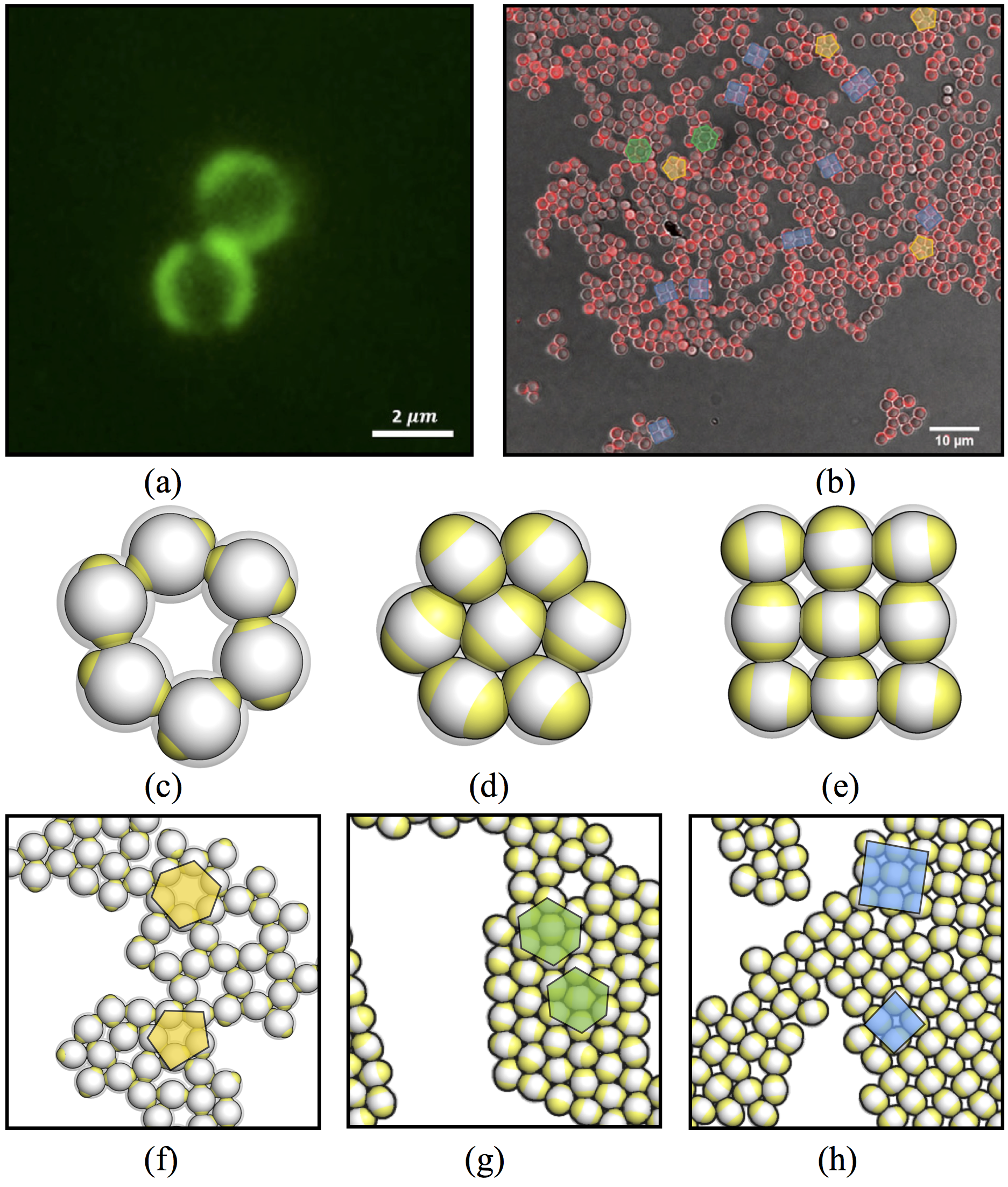}
\caption{Inverse patchy colloids (IPCs) with two positive polar patches and a negatively charged equatorial belt. (a) Confocal micrograph of two interacting silica IPCs.  (b) Overlaid brightfield and confocal micrograph of silica IPCs sedimented onto a glass surface where they form a gel-like structure. (c)-(e) Examples of a ring-like, hexagonal and square particle arrangement, respectively. (f)-(h) Simulation snapshots of IPCs under planar confinement: the particle arrangement depends on the patch size, the net particle charge and the substrate charge: (c) small patches, overall neutral particles and neutral substrate; (d) big patches, overall negative particles and neutral substrate, (e) big patches, neutral particles and positive substrate. Ring-like, hexagonal and square arrangements are highlighted in yellow, green and blue, respectively, in both experimental~\cite{peterjpcm_2015} and numerical~\cite{bianchi:2d2013,bianchi:2d2014} images. 
}
\label{fig:expsim}
\end{center} 
\end{figure}

If the system is confined to two dimensions and close to a homogeneously charged substrate, complex structures spontaneously emerge with well-defined translational and orientational order: depending on the charge ratio between the different entities involved, the investigated units can form surface layers with different densities (and possibly different responses to external stimuli) or do not assemble at all. 
The features of these assemblies also depend on other system parameters: disordered gel-like structures as well as crystalline domains forming an extended network (referred to as microcrystalline gel) were observed on varying the patch size~\cite{bianchi:2d2013}. The typical ordered particle arrangements can be characterized by grain- or flower-like patterns or by square-lattices, while in the emerging disordered structures the polar-equatorial bond induces ring-like structures involving typically up to five or six particles (see panel (c)-(h) of Figure~\ref{fig:expsim}). 
The same morphological features observed in simulations are found in experimental samples of IPCs sedimented on a glass substrate~\cite{peterjpcm_2015}. It is worth noting that, while in numerical samples all clusters have the same spatial and orientational order -- sharply defined by the patch size --, experimental IPCs form triangular, square and ring-like particle arrangements within the same sample  (see panel (b) of Figure~\ref{fig:expsim}). 
This difference is probably due to the patch size polydispersity of the IPCs synthesized so far.
It is worthwhile noting that even if the system is allowed to occupy a quasi two-dimensional space (with a slab height of a few particle diameters) most of the systems refuse to fully occupy the available space, but rather
form layers that extend parallel to the confining walls; this observation confirms the extraordinary strength of the intra-layer bonds, realized in one of the above mentioned bonding patterns~\cite{bianchi:2d2013,bianchi:2d2014}.
Moreover and most importantly, upon subtle changes of either the particle charge or the charge of the substrate (achieved, e.g., by pH changes or by changing the charge of the surface onto which the particles sediment), it is possible to reversibly switch the assembly process on and off as well as to induce a transformation from one specific spatial/orientational arrangement to another~\cite{bianchi:2d2014}. 
These results support the appropriateness of IPCs as model systems where the control over the ordering of the particles on substrates is feasible, so that colloidal monolayers with tunable properties can be easily realized. 

\begin{figure*}[htbp]
\begin{center} 
\includegraphics[width=\textwidth]{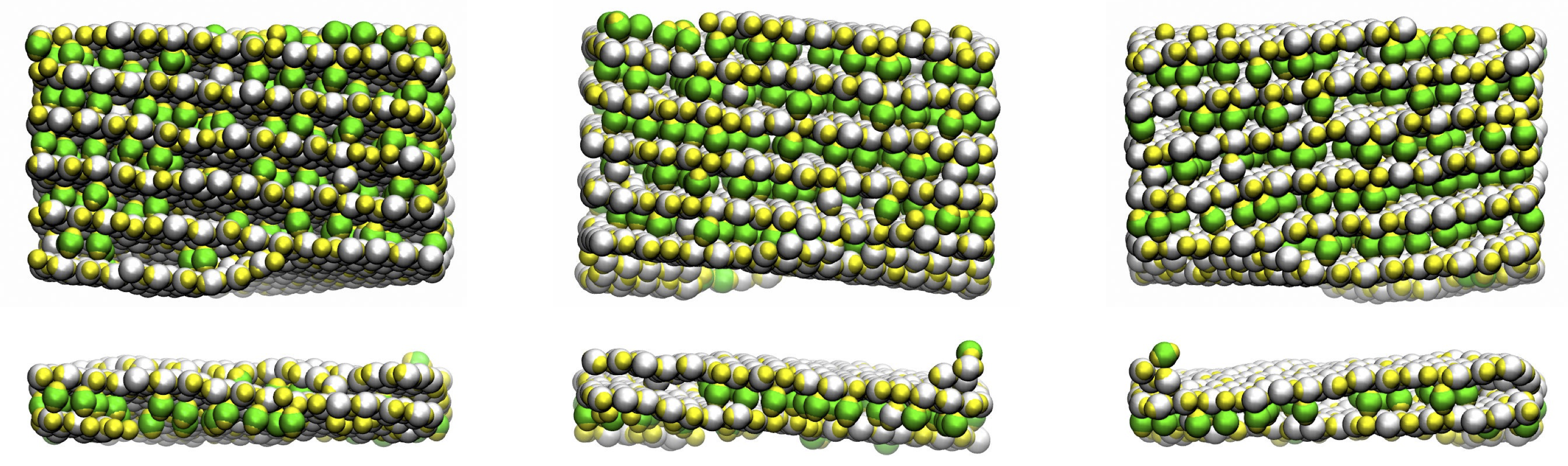}
\caption{Three different views of a self-assembled hybrid crystal-liquid phase: the layered phase is composed of parallel planes, where particles are arranged in a hexagonal, close-packed fashion (grey particles), and mobile inter-layer particles (green particles), arranged in a up-right orientation. The stoichiometric ratio between intra- and inter-layer particles is 2/7, while the distance between the planes is one particle diameter.}
\label{fig:3d}
\end{center} 
\end{figure*}

In the bulk, results accumulated up to now have only scratched the surface of the many possibilities offered by IPCs to materials design. The non-trivial interplay between attractive and repulsive directional interactions characterizes the fluid phase as well as the gas-liquid phase separation~\cite{silvanojpcm,Yura2015-2}, and gives rise to very interesting assembly and phase behaviors~\cite{ismene,evaemanJPCM,silvanonanoscale}. 
So far the tendency towards two-dimensional ordering is often maintained.
One option  for IPCs to stabilize a layered three-dimensional architecture is to pile up layers, forming thereby a lamellar crystal consisting of parallel, non close-packed monolayers (characterized internally by a close-packed, grain-like bonding pattern)~\cite{ismene}.  Even though particles from different layers are facing each other via their mutually repulsive equatorial regions, this lamellar structure is stable over a remarkably large range in parameter space: this stability is achieved as each layer experiences this repulsion from both sides, thus rendering the arrangement very stable.  The region of stability of this layered phase depends on the system parameters, e.g., it expands upon increasing the charge imbalance (i.e. on changing the pH of the solution) and/or upon reducing the interaction range (i.e. on changing the salt concentration of the colloidal suspension)~\cite{evaemanJPCM}.  An alternative strategy to form stable particle layers in three dimensions is realized in a lamellar phase which confines a relatively small number of intercalated particles, located between the monolayers (see Figure~\ref{fig:3d}). Here two parallel layers (each of them characterized by a close-packed grain-like intra-layer bonding pattern) are connected via inter-layer particles, oriented in a direction perpendicular to the adjacent layers~\cite{silvanonanoscale,silvanolong}. In  contrast to the previously observed lamellar phase (where the equilibrium distance between adjacent layers depends on the thermodynamic conditions), in this alternative lamellar architecture the equilibrium distance between adjacent planes and their mutual orientation is fixed by the inter-layer colloids. 
Again, the characteristic bonding pattern of IPCs -- the patches of the inter-layer particles form bonds with the equators of layer particles -- renders the structure highly stable over a remarkably large temperature range. Even at intermediate temperatures the lattice does not fully melt but transforms, instead, into an unconventional hybrid crystal-liquid structure: the inter-layer particles have sufficient energy to become mobile and start to diffuse between the layers which still maintain their stable and rigid internal structure. Eventually, as the temperature is further
increased, also the strong intra-layer bonds start to break up and the system melts, realizing thus the second step in a remarkable two-stage melting process. The additional amazing feature of the described hybrid crystal-liquid architecture is its ability to form spontaneously. The self-assembly of such a structure is guaranteed by (i) stable intra-layer bonds that stabilize the planar aggregates and (ii) strong inter-layer bonds that favor the stacking of the emerging planar assemblies. These two types of bonds together are responsible for the self-healing processes that are needed during the assembly. 

Lamellar phases represent just one of the many assembly scenarios offered by IPCs: preliminary investigations have shown that depending on the chosen parameters, IPCs with two identical patches can form an even wider zoo of exotic structures~\cite{Guenther_thesis_2012,silvano_phdthesis}, such as lattices percolated by mutually perpendicular sets of channels, that are reminiscent of zeolites. However, up to date their properties (ranging from the self-assembly processes to their actual physical properties) have not been explored nor have the possibilities offered by  IPCs with two different patches or with more than two patches have been taken under consideration.

\section{Conclusion and Outlook}\label{sec:conclusion_outlook}
Inverse patchy colloids (IPCs), i.e., colloidal particles with heterogeneously charged surfaces, represent a novel class of colloids with a high potential to self-assemble -- either spontaneously or triggered via external stimuli -- into complex and highly stable ordered particle arrangements. The advantage of this class of systems is two-fold: on one hand, IPCs have surface patterns that are simple enough to be experimentally accessible, yet complex enough to favor the stabilization of specific structures of interest; on the other hand, IPCs  offer the  opportunity of fine tuning the properties of the mesoscopic structures by means of external parameters, such as the pH and the salt concentration, either before or after the assembly process.

Current interest in the topic is highlighted by intensive research activity by a diverse selection of groups (both experimental and theoretical). The impressive progress that has been made in theoretical (mostly simulation based) investigations is complemented by the remarkable advancements in the synthesis of the particles. Nevertheless, the present status of research dedicated to the self-assembly capacities of IPCs and to the physical properties of the resulting structures still seems to be in its infancy and more interesting developments are to be expected. 

\section*{Acknowledgments}
We are indebted to G. Doppelbauer, S. Ferrari, M. Hejazifar, Y.V. Kalyuzhnyi, C. Niedermayer, E.G. Noya, E. Reimhult  and M. Stipsitz for their valuable contributions to the investigation of inverse patchy colloidal systems. 
The authors gratefully acknowledge financial support from the Austrian Science Fund (FWF) under Proj.~Nos. V249-N27 (EB), P27544-N28 (PvO), F41-SFB ViCoM (GK and CNL) and P23910-N16 (GK). 
Computing time on the Vienna Sicentific Cluster (VSC) is gratefully acknowledged.


\end{document}